\input lecproc.cmm
\input epsf.sty

\vsize=23true cm
\hsize=15true cm

\contribution{Unification of all blazars}
\contributionrunning{Unification of blazars}

\author{Gabriele Ghisellini} 
\authorrunning{G. Ghisellini}
\address{Osservatorio Astronomico di Brera, via Bianchi 46, Merate, Italy} 


\abstract{
The overall spectra (SED) of blazars, from radio to 
$\gamma$--ray energies, seem to obey well defined trends, with a 
continuity of properties between blazars of different classes.
To quantify this statement we can either investigate their $observed$ 
properties (see Fossati et al., this volume), or try to determine their
$intrinsic$ physical parameters by applying specific models, and trying
to fit their SED.
Results of the latter approach are reported here.
We applied simple, one--zone, homogeneous models to all blazars strongly 
detected in the $\gamma$--ray band, assuming or not the presence of seed 
(for the inverse Compton process) photons produced outside the active region.
Our results suggest that the SEDs of blazars are ruled by the amount of 
radiative cooling suffered by the electrons producing most of the emission.
In turn, the amount of cooling is ruled by the amount of the external
photon emission, which can be identified with radiation coming from the 
broad line clouds.
Blazar SEDs are therefore organized in a sequence:
objects with no or very weak emission lines (X--ray selected BL Lacs, or
HBL) are characterized by very high electron energies and a Compton
luminosity of the same order of the synchrotron one. 
These are TeV sources.
BL Lacertae objects selected in the radio band (or LBL) are characterized
by smaller electron energies, more total power and more line luminosity,
and by a larger ratio of the Compton to synchrotron luminosity.
They are GeV sources.
Increasing the total intrinsic power and the line emission luminosity, we
have smaller still electron energies (more cooling) and greater still
Compton to synchrotron power ratio. 
These are flat spectrum radio quasars, with a high energy peak located 
at MeV--GeV energies.
}

\titleb{1}{Introduction}

\noindent
Blazars come in many flavours: the basic distinction among them
is between lineless BL Lac objects and emission line flat spectrum
radio quasars (FSRQ). 
There are overlaps, since the division line of a rest frame equivalent width 
of 5 \AA~ of any emission line is crossed occasionaly by single objects.
Examples are PKS 0537--441 (usually a BL Lac) and 3C 279 (usually a FSRQ,
see Scarpa \& Falomo 1997).
BL Lac objects have been divided further into the subclasses of 
{\it high energy peak} (HBL) and {\it low energy peak} (LBL), according
to their overall synchrotron spectrum, which has a peak in the EUV--soft
X--ray band in the former objects and in the IR--optical band in the latter
(Giommi \& Padovani 1994).
FSRQ have been instead divided into the low polarization (LPQ) and high
polarization (HQP) subclasses, according if the level of optical
polarization is greater or smaller than 3\%.
HPQ are often identified with the {\it Optically Violent Variable} quasars 
(OVV), even if we know examples of LPQ which show rapid and strong 
variability.
Also here there are overlaps, since the degree of polarization is 
extremely variable, and the classification may reflect an observational 
bias (for objects observed more frequently there are more chances to observe 
an high degree of polarization).
In addition the contribution of the (probably thermal) blue bump component can 
dilute the non--thermal polarized emission, especially during faint states.

There is a general consensus upon the hypothesis that the extreme
phenomenology which characterizes the non--thermal emission of blazars is due 
to bulk motion of the emitting plasma, flowing in a collimated jet.
But differences in the bulk Lorentz factors $\Gamma$ or viewing angle
$\theta$, alone, are not sufficient to explain the different characteristics 
between different subclasses, such as BL Lac objects and FSRQ.
For instance, the idea that BL Lac objects are characterized by larger
$\Gamma$ and smaller $\theta$, to let the non--thermal continuum swamp
the lines and making the apparent superluminal velocities smaller
than in FSRQ, leads to predict an average larger luminosity for these
objects, contrary to what observed.
Furthermore, estimates of $<\Gamma>$ for BL Lacs and FSRQ indicate that
$<\Gamma>$ is larger in FSRQ (Madau Ghisellini \& Persic, 1987,
Ghisellini et al. 1993).

An attempt to unify HBL and LBL was made by Maraschi et al. (1986),
Ghisellini \& Maraschi (1989), Celotti et al. (1993),
based on the observational evidence that the X--ray luminosity of both 
subclasses is, on average, the same;
still, the X--ray surveys (which should therefore not be biased toward
one particular class) found systematically HBL type objects.
The idea was that the X--rays (thought to be synchrotron emission) were 
produced at the base of an accelerating jet, therefore in a zone where 
$\Gamma$ is small, resulting in an emission more isotropically distributed 
that the optical or radio (coming further out along the jet).
Therefore there should be a larger solid angle available for detecting BL Lacs 
in X--rays, the majority of which should be characterized by not strongly 
Doppler enhanced optical and radio emission, as observed.
Unfortunately, this idea does not account properly for the (large) shift in 
the energy of the synchrotron peak, nor for the difference recently found 
between X--ray spectra: HBL tend to have steeper X--ray spectral 
indices than LBL, indicating that in the former class we see the steep tail 
of the synchrotron emission, while in the latter objects we see the flatter 
Compton component  (see e.g., Comastri et al. 1995, 1997).
Another evidence against the ``accelerating jet model" is the variable and 
strong $\gamma$--ray emission detected in some BL Lacs: for the $\gamma$--ray 
to escape without suffering strong $\gamma$--$\gamma$ absorption, we need 
large $\Gamma$ even in the inner, X--ray emitting, jet (Dondi \& Ghisellini,
1995).

Since blazar emission is beamed, there must be many sources, whose jet is 
pointed far from the line of sight, observed to have very different properties.
These sources form the so--called ``parent population" of blazars, whose
identification is still under debate.
The current understanding, after the work of Urry \& Shafer (1984),
Padovani \& Urry (1992), Celotti et al. (1993), Maraschi \& Rovetti (1994) 
is that the BL Lacs `live' in FRI radio--galaxies, while the more powerful 
FRII should host FSRQ (for a review, see Urry \& Padovani 1995). 

\titleb{2}{Gamma--loud blazars}

The $\sim$60 blazars detected so far by EGRET (Fichtel 1994; von Montigny
et al. 1995; Thompson et al. 1995) lead us to believe that the
$\gamma$--ray emission is a characteristic of the entire blazar class.
Blazars detected by EGRET (and by Cherenkov detectors, such as WHIPPLE and 
HEGRA, in the TeV band) include objects belonging to all different subclasses 
of blazars.

Only now we have the complete view of the overall emission of blazars, 
and for most sources only now we know in what band the 90\% of the luminosity
is emitted.
Now we can model their SED for deriving the intrinsic physical parameters of
the $\gamma$--ray emitting region, which dominates the bolometric output.

One is then driven to ask if the different subclasses in which we classify 
blazars are the result of some deeper physical distinction among them.
In particular, there may exist one (or more) changing parameter 
determining the different properties (and classification) of blazars.

To investigate this problem, one can study the overall observed 
spectral properties of (relatively large) sample of blazars, 
to search for trends and/or correlations. 
This approach has been adopted by Fossati et al. 1997 (see also
Fossati et al. this volume).
Alternatively, one can try to model their emission in order to find
the {\it intrinsic} values of the physical parameters, once beaming
and relativistic effects are properly taken into account.
This is the approach adopted in Ghisellini et al. (1997),
whose results will be discussed here.

The first problem to face with this approach is the choice of the model,
among the many already put forward to explain the entire SED of blazars.
In the model of Mannheim (1993) shock--accelerated 
electrons and protons originate two different populations of emitting particles
(electrons and electron--positron pairs), responsible of the entire
SED of blazars by emitting synchrotron photons.
In other models, instead, a single populations of electrons emits
far IR (or even radio) to UV--soft X--ray radiation by synchrotron, and higher
frequencies by the Inverse Compton process.
Models of this kind differ by the adopted geometry (one--zone homogeneous
models or inhomogeneous jet models), and by the nature of the target photons
to be upscattered in energy by the Inverse Compton process.
These photons could be synchrotron photons (Maraschi, Ghisellini,
\& Celotti, 1992; Bloom \& Marscher, 1993)
photons produced in the accretion disk (Dermer \& Schlickeiser, 1993),
or in the broad line region (BLR) illuminated by the disk
(Sikora, Begelman \& Rees, 1993, Blandford \& Levinson,
1995), or self--illuminated by the jet (Ghisellini \& Madau, 1996).
Finally, target photons could be produced by a dusty torus surrounding the
blazar nucleus (Wagner et al. 1995).
These models have been applied to specific sources, and often more than one 
model could fit the same data (see von Montigny et al. 1997 for 3C 273, 
Ghisellini, Maraschi \& Dondi 1996 for 3C 279, Comastri et al. 1997 for
0836+710).

Ghisellini et al. (1997) have chosen to apply two one--zone and homogeneous
models: the synchrotron self--Compton (SSC) model and the 
``external Compton" model (EC), in which the main contribution to the target 
photons is produced outside the $\gamma$--ray production region, by an 
(yet) unspecified mean.
Data were collected from the literature for all the 45 blazars for which we 
have informations on their $\gamma$--ray spectral shape (to locate the energy
of the peak of the high energy emission) and on their redshift (to measure 
the apparent luminosities).
Among them there are 12 BL Lacs (4 HBL, 8 LBL), 15 LPQ, 15 HPQ and 3 FSRQ 
for which we do not have yet polarization measurements.

The model assumes that the emission region is a sphere of radius $R$,
moving with a bulk Lorentz factor $\Gamma$ at an angle $\theta\sim
1/\Gamma$ toward the observer. 
The magnetic field $B$ is tangled and uniform, and the particle distribution
is found solving the continuity equation, balancing continuous injection
and radiative cooling.
Pair production and Klein Nishina effects are taken into account.
Electrons are injected with a power law distribution ($\propto \gamma^{-s}$)
between $\gamma_{min}$ and $\gamma_{max}$.
In the simple case of Compton scattering in the Thomson regime and
no pair production, the equilibrium distribution has a broken power law
shape $\propto \gamma^{-2}$ for $\gamma<\gamma_b=\gamma_{min}$,
and $\propto \gamma^{-(s+1)}$ above.
$\gamma_b$ is a crucial parameter, since it corresponds to the energy
of those electrons responsible of both the synchrotron and the Compton peak.
In the EC case, we further assume that the source is embedded in a photon
bath of radiation energy density $U^\prime_{ext}$
(in the comoving frame), distributed in energy as a blackbody peaked at 
$\nu^\prime_0\sim 10^{16}$ Hz.

\titleb{4}{Results of fitting the SED of blazars}

Fig. 1 shows some examples of the SED of BL Lacs and FSRQ, together with 
the fitting models. 
Note that, on the basis of the fit, one cannot discriminate between
the SSC and the EC model.
However, in the SSC case, we inevitably find large values of $\Gamma$,
small values of the magnetic field and small synchrotron (intrinsic) powers.
This implies that externally produced photons, not to contribute to
the Compton process, must have a very small luminosity.
This may be the case for $some$ BL Lac, but it is certainly not for FSRQ,
whose luminosity in the broad lines exceeds the found limit.
Therefore in the following we concentrate on the discussion of the
results of the EC fit for all sources.
For $some$ BL Lacs (e.g. 0235+164, 0537--441, 1604+159), some amount of 
external photons is indeed required to obtain a better fit, while for
the remaining this can be taken as an upper limit not to worsen the fit.

\topinsert
\vglue -0.6 cm
\centerline{\epsfxsize=14 cm \epsfbox{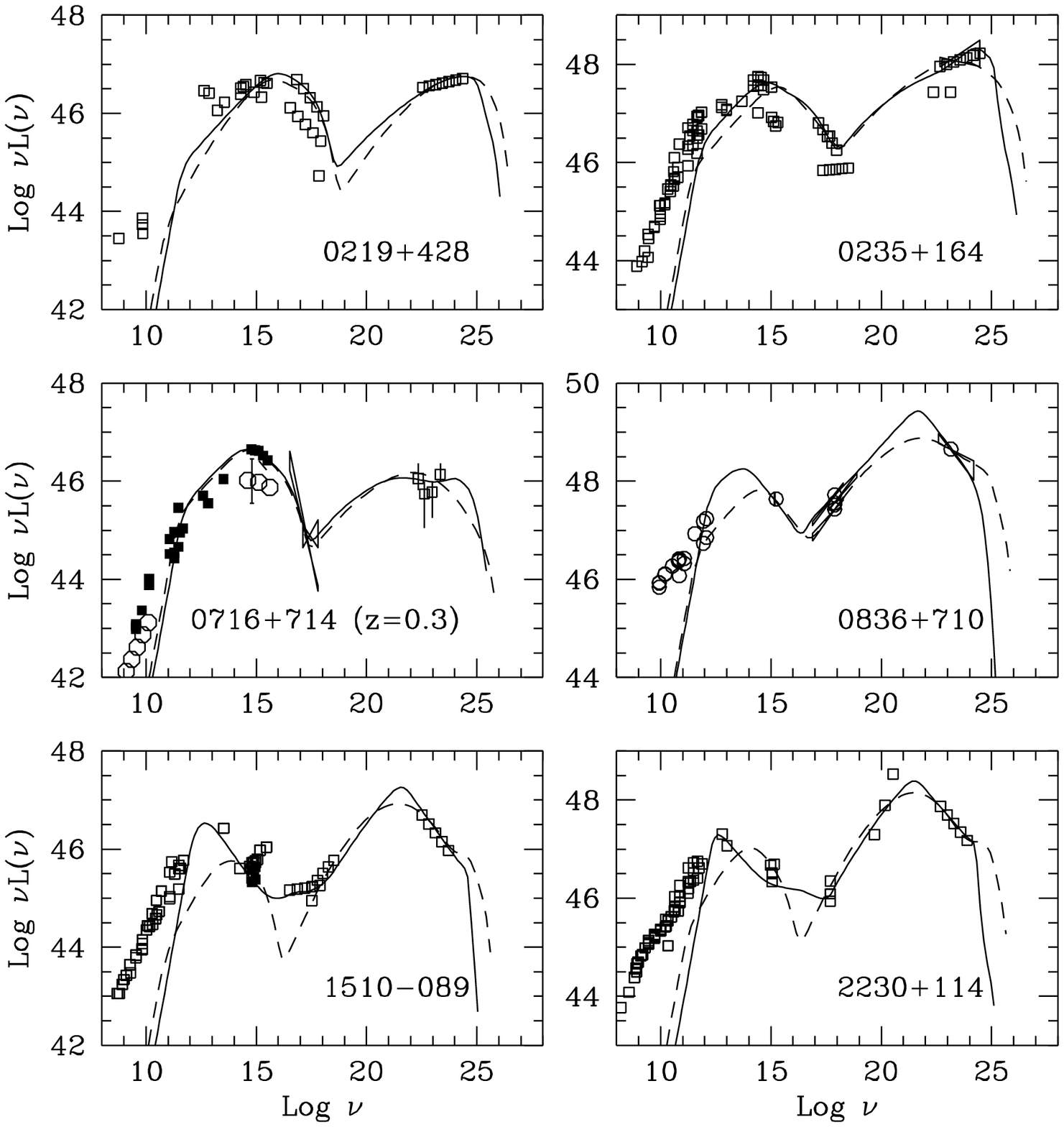}}
\vskip -1.5 true cm
\setbox0=\vtop{
{\figure{1}{Selected examples of SEDs of blazars, fitted by the SSC
model (dashed line) and the EC model (solid line). 0219+428, 0235+164
and 0716+714 are BL Lac objects, while 0836+710, 1510--089 and 2230+114
are FSRQ. Adopted from Ghisellini et al. 1997.}}
}
\line{\box0}
\endinsert

\titlec{4.1}{Average values of the parameters}

We require that the beaming factor (hence $\Gamma$) does not
exceed a value of $\sim 20$, to be consistent with values of the
observed superluminal velocities.
The size $R$ is constrained by the observed variability timescales,
required not to exceed $\sim$1 day.
Furthermore, the compactness corresponding to the injected power
$\ell_{inj}=L_{inj}\sigma_T/(Rmc^3)$ is limited to values roughly
less than unity, to avoid strong pair production.

With these constraints, we find that, on average, the size $R$
is of the order of $10^{16}$--$10^{17}$ cm, the beaming factor
$\delta\sim 15$, the magnetic field $B\sim 1$ Gauss and the injected
compactness $\ell_{inj}\sim 0.1$.
The slope of the injected electron distribution is steep ($s\sim$2--3), 
making $\gamma_{max}$ energetically unimportant.
The most relevant parameter is $\gamma_b$ (equal to $\gamma_{min}$,
the miminum $\gamma$ of the {\it injected} distribution), because it 
controls the locations of the synchrotron and Compton peaks.
BL Lacs are characterized by larger values of $\gamma_{b}$ than FSRQ.

\titlec{4.2}{Correlations}

Fig. 2a shows the correlation between the injected power and the amount of 
external photons (measured by an `effective compactness' $\ell_{ext}$: 
it is defined as if these photons were produced within the blob itself. 
In this way we can directly compare $\ell_{inj}$ and $\ell_{ext}$).
As can be seen, there is a strong correlation between the two
(intrinsic) compactnesses. 
As expected, BL Lac objects lie on a separate portion of the plane:
some LBL lie on the extrapolation of the correlation defined by
FSRQ, while for HBL $\ell_{ext}$ is smaller.

\topinsert
\vglue -0.6 cm
\vskip 0.5 true cm
\centerline{\epsfxsize=13.5 cm \epsfbox{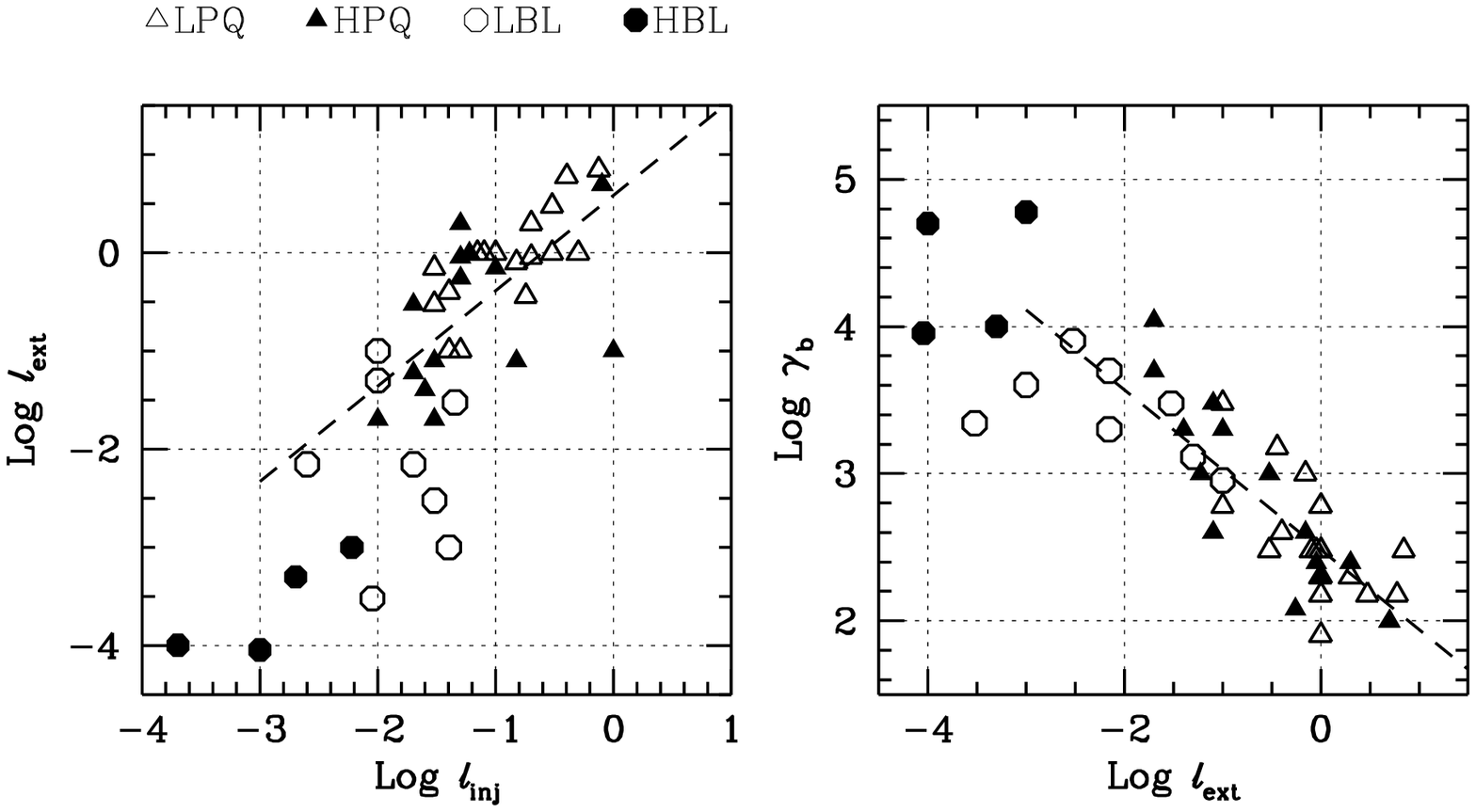}}
\vskip -9 true cm
\setbox0=\vtop{
{\figure{2}{{\it Left panel:} Correlation of the compactness $\ell_{ext}$ 
of external photons and the compactness injected in the source in the form 
of relativistic electrons. Both are intrinsic (comoving) quantities.
{\it Right panel:} Correlation between the energy $\gamma_b$ of the
break in the electron distribution and $\ell_{ext}$.
Dashed lines are fits to the FSRQ only.
}}
}
\line{\box0}
\endinsert

\topinsert
\vglue -0.6 cm
\centerline{\epsfxsize=13.5 cm \epsfbox{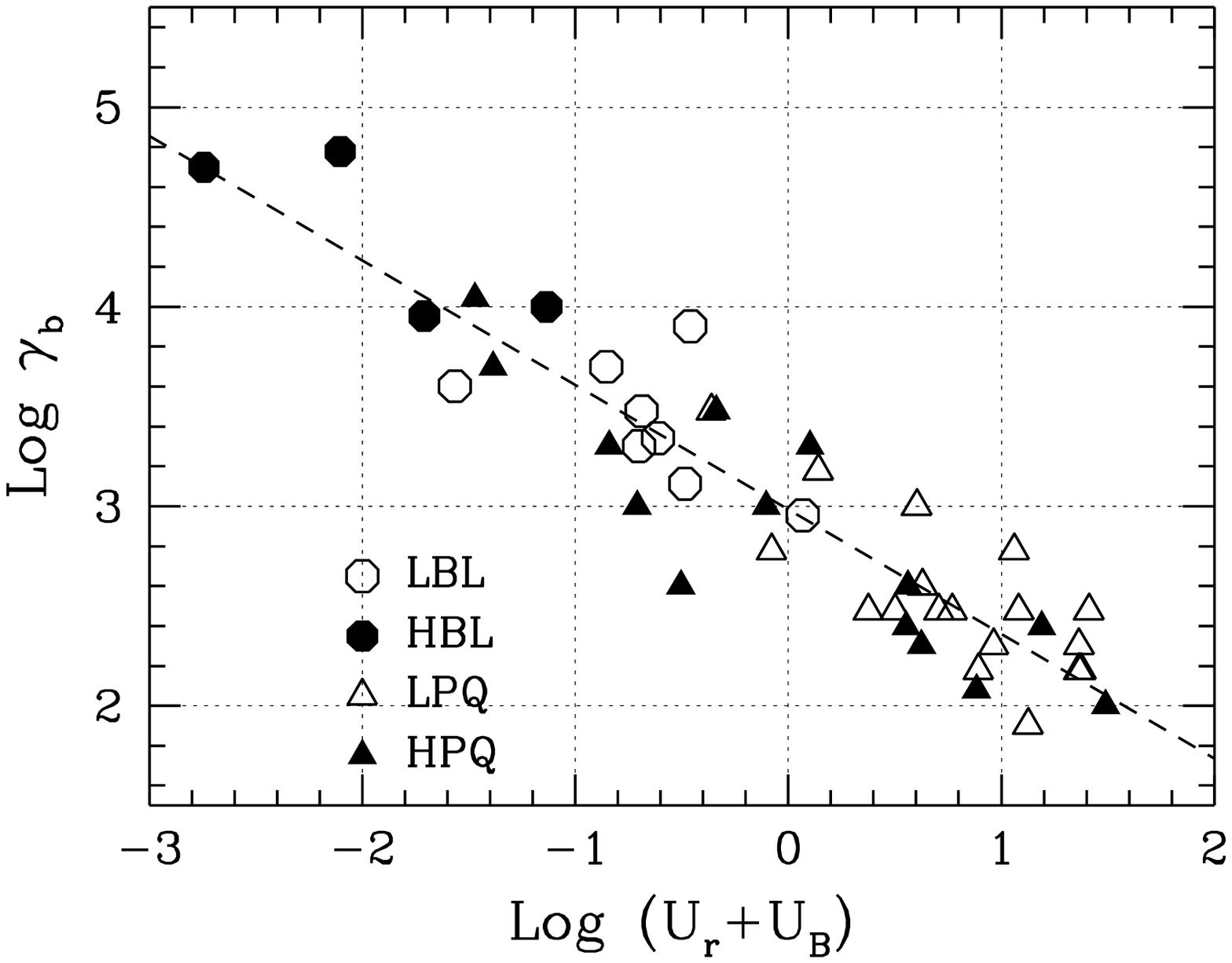}}
\vskip -6 true cm
\setbox0=\vtop{
{\figure{3}{Correlations between the value of the Lorentz
factor of the electrons emitting at the peaks of the SED
and the total energy density (radiative+magnetic). }}
}
\line{\box0}
\endinsert

Fig. 2b shows the correlation between $\gamma_b$ and $\ell_{ext}$.
Also in this case there is a strong correlation: large values
of $\ell_{ext}$ imply small value of $\gamma_b$, hence a 
synchrotron and Compton spectrum peaking at lower energies.
Again, BL Lacs occupy a separate portion of the diagram,
with HBL at one extreme.

Fig. 3 shows the correlations betwen $\gamma_b$ and the total
(intrinsic) amount of energy density, including radiation [both
internal (produced by synchrotron) and external], and magnetic field.
This is the strongest correlation we found.
A linear correlation analysis (of the logarithm of the two
quantities) yields $\gamma_b\propto U^{-0.6}$ and is shown by
the dashed line.

Other important correlations found concern the `Compton dominance'
parameter $L_c/L_s$ (ratio of the Compton to synchrotron luminosity)
and either $\gamma_b$, or $\ell_{inj}$, or $\ell_{ext}$, in the
sense that the $L_c/L_s$ increases for larger $\ell_{inj}$ or $\ell_{ext}$,
and for smaller values of $\gamma_b$.

\titlec{4.3}{Cooling vs acceleration}

We can investigate further the found correlation between $\gamma_b$ and $U$.
Assume in fact that the (yet unknown) acceleration mechanism 
accelerates electrons up to some energy $\gamma$, where the cooling
rate becomes competitive with acceleration.
At this energy we then have $\dot\gamma_{cool}\sim \dot\gamma_{acc}$.
Since $\dot\gamma_{cool}\propto \gamma^2 (U_r+U_B)$, we have
$\gamma_b = {\rm const}\times \,  (\dot\gamma_{acc}/ U)^{1/2}$.
We then have (approximately) the correlation we found
{\it if $\dot\gamma_{acc}$ is roughly constant}.

\titleb{5}{The blazar sequence}

Our results indicate that all blazars can be organized in a well
defined sequence, according to their intrinsic power, or the power
in the external radiation field, as produced by the broad line clouds.
This power increases from HBL to LPQ, which are at the extremes
of the sequence. 
At the same time, the energy of the most relevant emitting electrons 
decreases, and the Compton dominance increases.

We then propose this scenario:

$\bullet$ HBL are characterized by the smallest intrinsic power, and by the 
weakest lines (and/or external radiation).
Suffering less cooling, electrons can be accelerated at very high
energies, enough to produce TeV emission by self--Compton emission,
while their synchrotron spectrum peaks in the soft (or even hard) 
X--ray band.
Their  Compton dominance is at the low end of the blazar distribution.

$\bullet$ LBL are more powerful, and in some of them the external
radiation field could be important.
Radiative cooling is more severe, and electrons can attain smaller
energies than in HBL.
Correspondingly, the synchrotron peaks in the optical and the Compton
emission peaks in the GeV band.
Due to the increased (Compton) cooling, their Compton dominance
is larger than in HBL.

$\bullet$ FSRQ are more powerful still, and their external radiation field
is even more important, inducing a great radiative cooling.
In this situation electrons can attain relatively
small energies, and their synchrotron 
emission peaks in the IR, while the (dominating) Compton emission peaks
in the MeV or in the MeV--GeV band.

\begrefchapter{References}

\ref Blandford, R.D. \& Levinson, A. 1995, ApJ, 441, 79

\ref Bloom, S.D. \& Marscher, A.P., 1993, in Proceeedings of the Compton
Symposium, eds. M. Friedlander \& N. Gehrels (New York: AIP), 578

\ref Celotti A., Maraschi L., Ghisellini G., Caccianiga A.,
Maccacaro T.,  1993, ApJ, 416, 118

\ref Comastri, A., Molendi, S. \& Ghisellini, G., 1995, MNRAS, 277, 297

\ref Comastri, A., Fossati, G., Ghisellini, G., \& Molendi, S., 1997, 
ApJ, 480, 534 

\ref Dermer, C. \& Schlickeiser, R., 1993, ApJ, 416, 458

\ref Dondi, L. \& Ghisellini, G., 1995, MNRAS, 273, 583

\ref Fichtel, C.E. et al., 1994, ApJS 94, 551

\ref Fossati, G., Celotti, A., Ghisellini, G., \& Maraschi L., 1997a, 
M.N.R.A.S., in press.

\ref Fossati, G., Celotti, A., Comastri, A., Ghisellini, G., \& Maraschi L., 
1997b, in preparation

\ref Ghisellini, G., Celotti, A., Comastri, A., Fossati, G., Maraschi, L., 
1997, in preparation

\ref Ghisellini, G., \& Maraschi, L., 1989, ApJ, 340, 181

\ref Ghisellini, G., 1989, MNRAS, 238, 449

\ref Ghisellini, G., Padovani, P., Celotti, A. \& Maraschi, L., 1993, ApJ, 
401, 65

\ref Ghisellini, G.  \& Madau, P., 1996, MNRAS, 280, 67

\ref Ghisellini, G., Maraschi, L. \& Dondi, L. 1996 

\ref Ghisellini, G., Celotti, A., Fossati, G. \& Comastri A., 1997. in prep.

\ref Giommi, P., Padovani, P., 1994, MNRAS, 268, L51

\ref Madau, P., Ghisellini, G., \& Persic, M., 1987, MNRAS, 224, 257

\ref Mannheim, K., 1993, A\&A, 269, 67

\ref Maraschi, L., Ghisellini, G. \& Celotti, A., 1992, ApJ, 397, L5

\ref Maraschi L., Ghisellini, G., Tanzi, E.G., \& Treves, A. 1986, 
ApJ, 310, 325.

\ref Maraschi L. \& Rovetti F., 1994, ApJ, 436, 79

\ref Padovani, P. \& Urry, C.M. 1992, ApJ, 387, 449 

\ref Scarpa, R. \& Falomo, R., 1997, A\&A in press

\ref Sikora, M., Begelman, M.C. \& Rees, M.J. 1994, ApJ, 421, 153

\ref Thompson D.J. et al. 1995, ApJS, 101, 259

\ref Urry, C.M. \& Padovani, P., 1995, PASP, 107, 803

\ref Urry, C.M. \& Shafer, R.A., 1984, ApJ 280, 569

\ref von Montigny, C., et al., 1995, ApJ, 440, 525

\ref von Montigny, C. et al., 1997, ApJ, in press.

\ref Wagner, S.J., Camenzind, M., Dreissigacker, O., et al., 1995, 
A\&A 298, 688

\endref

\bye\bye